\documentclass[english,aps,prl,reprint,superscriptaddress,nofootinbib,nobibnotes,notitlepage,preprintnumbers,showpacs]{revtex4-1}
\pdfoutput=1
\usepackage{lmodern}
\DeclareMathAlphabet{\mathbbold}{U}{bbold}{m}{n} 

\usepackage[T1]{fontenc}
\usepackage[latin9]{inputenc}
\usepackage{geometry}
\usepackage{color}
\usepackage{babel}
\usepackage{amsmath}
\usepackage{amssymb}
\usepackage{cancel}
\usepackage{bbold}
\usepackage{graphicx}
\usepackage{esint}
\usepackage[unicode=true,pdfusetitle,
 bookmarks=true,bookmarksnumbered=false,bookmarksopen=false,
 breaklinks=false,pdfborder={0 0 1},backref=false,colorlinks=true]
 {hyperref}
\hypersetup{
 citecolor=blue,linkcolor=blue,urlcolor=blue}

\makeatletter

 \@ifundefined{textcolor}{}
 {%
   \definecolor{BLACK}{gray}{0}
   \definecolor{WHITE}{gray}{1}
   \definecolor{RED}{rgb}{1,0,0}
   \definecolor{GREEN}{rgb}{0,1,0}
   \definecolor{BLUE}{rgb}{0,0,1}
   \definecolor{CYAN}{cmyk}{1,0,0,0}
   \definecolor{MAGENTA}{cmyk}{0,1,0,0}
   \definecolor{YELLOW}{cmyk}{0,0,1,0}
 }

\usepackage{babel}

\makeatletter
\def\simgt{\mathrel{\lower2.5pt\vbox{\lineskip=0pt\baselineskip=0pt
           \hbox{$>$}\hbox{$\sim$}}}}
\def\simlt{\mathrel{\lower2.5pt\vbox{\lineskip=0pt\baselineskip=0pt
           \hbox{$<$}\hbox{$\sim$}}}}
\makeatother

\newcommand{\be}{\begin{equation}}
\newcommand{\ee}{\end{equation}}
\newcommand{\bea}{\begin{eqnarray}}
\newcommand{\eea}{\end{eqnarray}}
\newcommand{\Ref}[1]{Ref.~\cite{#1}}
\newcommand{\Fig}[1]{Fig.~\ref{#1}}
\newcommand{\Eq}[1]{Eq.~(\ref{#1})}

\newcommand{\ket}[1]{|#1\rangle}

\begin{document}

\preprint{\hbox{CALT-TH-2015-033} }

\title{Splitting Spacetime and Cloning Qubits: \\Linking No-Go Theorems across the ER=EPR Duality
}

\author{Ning Bao}
\affiliation{Institute for Quantum Information and Matter}
\affiliation{Walter Burke Institute for Theoretical Physics, \\California Institute of Technology, Pasadena, CA 91125, USA}
\author{Jason Pollack}
\affiliation{Walter Burke Institute for Theoretical Physics, \\California Institute of Technology, Pasadena, CA 91125, USA}
\author{Grant N. Remmen}
\affiliation{Walter Burke Institute for Theoretical Physics, \\California Institute of Technology, Pasadena, CA 91125, USA}
\email{ningbao@theory.caltech.edu,\\jpollack@theory.caltech.edu,\\gremmen@theory.caltech.edu}

\begin{abstract}
We analyze the no-cloning theorem in quantum mechanics through the lens of the proposed ER=EPR (Einstein-Rosen = Einstein-Podolsky-Rosen) duality between entanglement and wormholes. In particular, we find that the no-cloning theorem is dual on the gravity side to the no-go theorem for topology change, violating the axioms of which allows for wormhole stabilization and causality violation. Such a duality between important no-go theorems elucidates the proposed connection between spacetime geometry and quantum entanglement.
\end{abstract}

\maketitle

\section{Introduction}\label{sec:Intro}

The connection between entanglement and geometry is an unexpected stepping-stone on the path to an understanding of quantum gravity. Historically originating from black hole thermodynamics \cite{BHLaws,Bekenstein} and later in the context of the holographic principle \cite{Holography1,Holography2}, the AdS/CFT correspondence \cite{AdSCFT,Witten,MAGOO}, entropy bounds \cite{Bousso}, and the Ryu--Takayanagi formula \cite{RT}, the relation between quantum entanglement and spacetime geometry is increasingly thought to be an important feature of a consistent theory of quantum gravity. Underscoring this view is recent work on deriving the Einstein equations holographically from entanglement constraints \cite{Faulkner} and perhaps even spacetime itself from qubits \cite{Swingle,AdSMERA}. However, significant puzzles remain. The classic black hole information paradox \cite{HawkingInfo,Mirrors} has given way to new questions about black hole interiors and their entanglement with Hawking radiation \cite{AMPS,Braunstein}. One of the most drastic, albeit promising, proposals to arise from these debates is the so-called ER=EPR duality \cite{ER=EPR}.

The ER=EPR correspondence \cite{ER=EPR} is a compelling~\cite{JensenKarch,Sonner} proposal for an exact duality between Einstein-Podolsky-Rosen (EPR) pairs \cite{EPR}, that is, qubits entangled in a Bell state \cite{Bell}, and nontraversable wormholes, that is, Einstein-Rosen (ER) bridges \cite{ER,Kruskal,Fuller}. More specifically, the ER=EPR proposal generalizes the notion of entangled black hole pairs at opposite ends of an ER bridge, by asserting that every pair of entangled qubits is connected by a Planck-scale quantum wormhole. The proposal, if true, would have profound implications for AdS/CFT and suggest a solution to the firewall paradox of \Ref{AMPS}, not to mention the fundamental shift it would induce in our understanding of both quantum mechanics and general relativity. 

The ER=EPR correspondence might allow the exploration of gravitational analogues of fundamental properties of quantum systems (and vice versa). In particular, we can check whether there is a precise correspondence between no-go theorems in quantum mechanics and similar no-go theorems in gravity. Arguably the most celebrated no-go theorem in quantum mechanics is the no-cloning theorem \cite{WZ}, which prohibits the duplication of quantum states. 

In this paper, we investigate the manifestation of the no-cloning theorem on the gravitational side of the ER=EPR duality. In particular, we show that violation of the no-cloning theorem is dual under ER=EPR to topology-changing processes in general relativity, which, via classical topology-conservation theorems \cite{GerochThesis,Centenary,Tipler,Tipler2,
Hajicek,TopoCensor,TopoCensor2}, lead to causal anomalies through violation of the Hausdorff condition (which leads to the breakdown of strong causality), creation of closed timelike curves (CTCs), or  violation of the null energy condition (NEC) (which allows for wormhole traversability and hence CTCs). While the validity of ER=EPR requires both unitarity and wormhole nontraversability, it is interesting that these two requirements seem to be fundamentally related: the no-cloning theorem and the topology-conservation theorem, both of which are related to causality, are in fact dual no-go theorems under ER=EPR.

\section{Quantum Cloning}\label{sec:Cloning}
Here, we reconstruct the standard argument for why the no-cloning theorem prohibits superluminal signaling~\cite{Dieks}. Assume that cloning of states is allowed, that is, that there exists an operation that takes an arbitrary state $\ket{\Psi}$ in a product state with some $\ket{0}$ state and replaces the $\ket{0}$ state with $\ket{\Psi}$:
\begin{equation}
\ket{\Psi}_A\ket{0}_B \rightarrow \ket{\Psi}_A\ket{\Psi}_B.\label{eq:clone}
\end{equation}
Suppose that there exists an EPR spin pair, the state $(\ket{00}+\ket{11})/\sqrt{2}$. We give one spin to each of a pair of individuals, Alice and Bob, who may then move to arbitrary spacelike separation. Alice now makes a decision as to the classical bit she wishes to communicate: to send a ``1'', she measures in the $\sigma_z$ basis, while to send a ``0'', she does nothing. 

Bob now proceeds to clone his qubit as in \Eq{eq:clone}. Note that each of his cloned qubits remains maximally entangled with Alice's qubit, in violation of monogamy of entanglement, while remaining unentangled with each other. By measuring enough of his own qubits in the $\sigma_z$ basis, Bob can determine, to any desired degree of confidence, whether Alice performed a measurement or not: his measurements will all yield the same result if Alice performed a measurement, but will be equally and randomly split between the two outcomes if she did not.  As this experiment does not depend on their separation, Bob's utilization of cloning and their shared entanglement has allowed Alice to send one classical bit to Bob acausally.

\section{Black Hole Cloning}\label{sec:BH}

In order to geometrically interpret the no-cloning theorem using the ER=EPR proposal, we need a system with both a high level of entanglement (like the EPR pair just considered) and a robust geometric description.
One such system is the eternal AdS-Schwarzchild black hole, which is described in AdS/CFT by two noninteracting large-$N$ CFTs in a thermally entangled state on the boundary sphere \cite{Maldacena:2001kr,VanRaamsdonk:2010pw}:
\begin{equation}
\ket{\Psi} = \frac{1}{\sqrt{Z}}\sum_n e^{-\beta E_n/2}\ket{n}_L\otimes\ket{n}_R,\label{eq:2CFTs}
\end{equation}
where $\ket{n}_{L (R)}$ is the $n$th eigenstate on the left (respectively, right) CFT with energy $E_n$, $\beta$ is the inverse temperature, and $Z$ is the partition function. In this state, the reduced density matrices $\rho_{L,R}$ of either side are identically thermal.
If both exterior regions of the geometry are considered \cite{Horowitz:1998xk,Maldacena:2001kr,VanRaamsdonk:2010pw}, this state describes a spacetime consisting of two separate AdS-Schwarzchild regions that are spatially disconnected outside the horizon but linked by an ER bridge between a maximally entangled\footnote{Strictly speaking, the state is only truly maximally entangled when $\rho_L=\rho_R=\mathbbold{1}$, i.e., when $\beta\rightarrow 0$, but we adopt the terminological abuse of \Ref{ER=EPR}.} pair of black holes with temperature $\beta^{-1}$.
This is a concrete realization of ER=EPR: to reiterate, the two black holes are both maximally entangled~(EPR) and connected by a nontraversable wormhole~(ER). 
It will be convenient to consider the slight generalization of this setup in which the two black holes share the same asymptotic space.
As discussed in \Ref{ER=EPR}, such black hole pairs can be naturally obtained as an instanton solution in a geometry with a constant magnetic field.

\begin{figure}[t]
\begin{center}
\includegraphics[width=.47\textwidth]{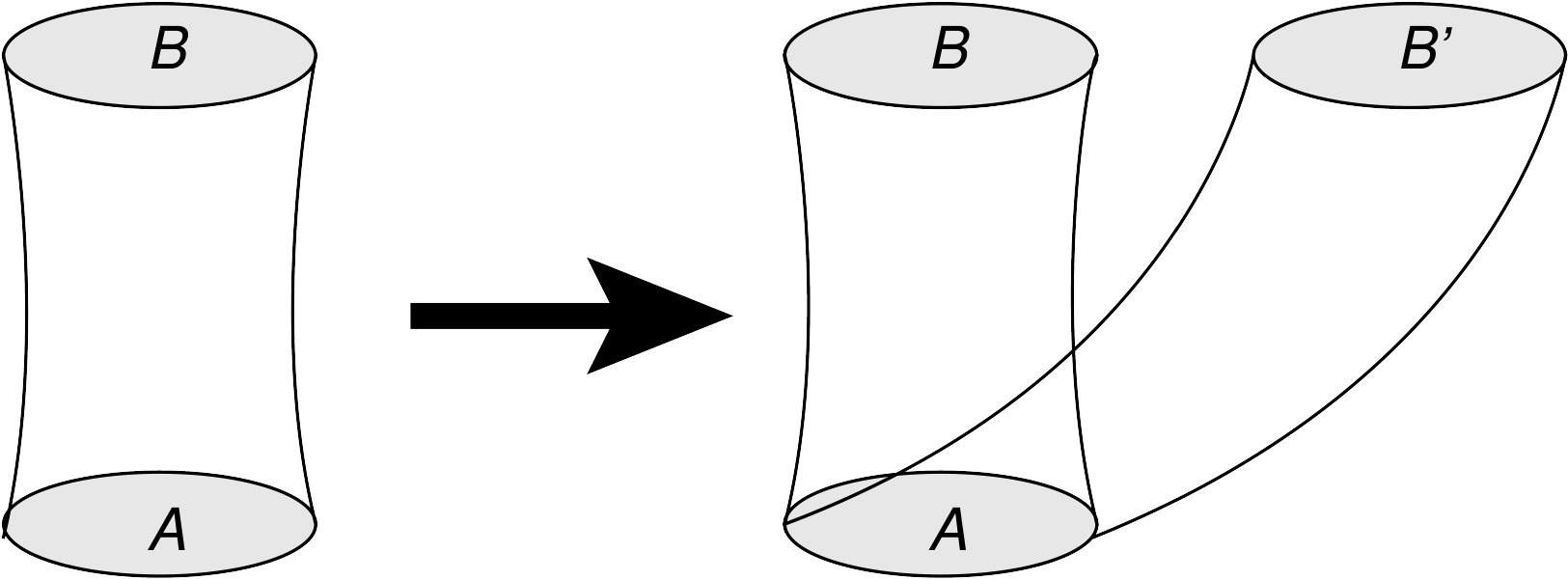}
\end{center}
\vspace*{-5mm}
\caption{Illustration of the black hole cloning thought experiment in the context of the ER=EPR conjecture. If Bob has access to a device that can clone quantum states, he can transform black hole $B$, which is entangled with $A$, into two black holes $B$ and $B'$, each connected to $A$ via an ER bridge.}
\label{fig:CloningWormholes} 
\end{figure}

We now consider repeating the experiment in the previous section using entangled black holes instead of qubits, as depicted in \Fig{fig:CloningWormholes}.
Alice and Bob, who live in an asymptotically-AdS spacetime, are each given access to a Schwarzschild black hole, with the two black holes, labeled $A$ and $B$ respectively, maximally entangled and therefore connected by a nontraversable wormhole. 
If Bob now clones all the degrees of freedom on his stretched horizon \cite{Complementarity}, he is left with two black holes $B$ and $B'$, each of which is connected by an ER bridge to Alice's black hole. That is, cloning is dual to change of spacetime topology under ER=EPR. 

\begin{figure}[t]
\begin{center}
\includegraphics[width=.47\textwidth]{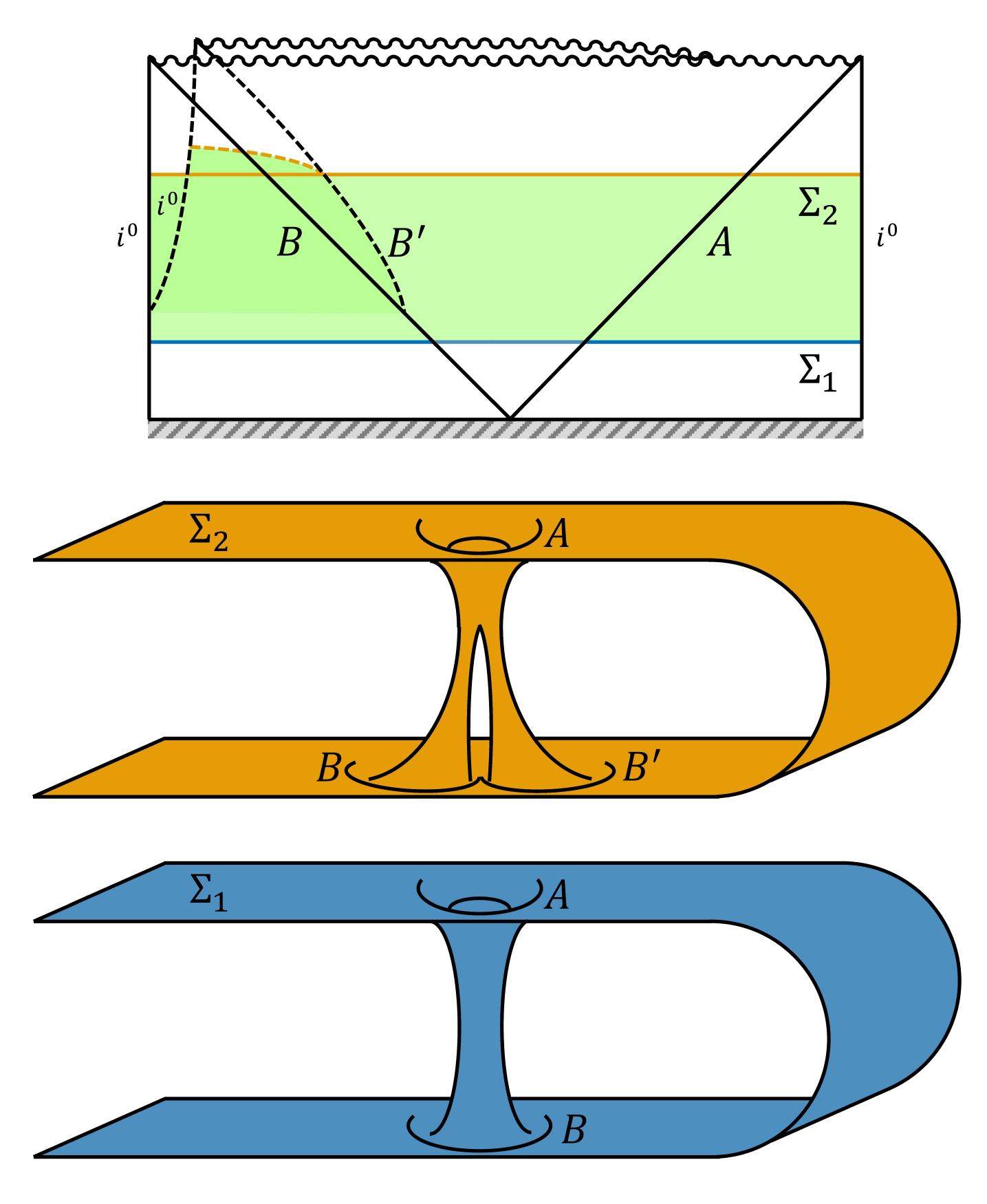}
\end{center}
\vspace*{-5mm}
\caption{Penrose diagram for the topology-change process depicted in \Fig{fig:CloningWormholes}, with spatial slices $\Sigma_1$ (blue) and $\Sigma_2$ (orange) shown as embedding diagrams. The spacetime region $M$ (green) is indicated; the compact region $K$ with nontrivial topology is bounded by horizons $A$, $B$, and $B'$. All of the spatial infinities $i^0$ are identified, as the black holes share the same asymptotically-AdS spacetime. The diagonal stripes at the bottom of the Penrose diagram indicate that the half of the spacetime containing the past horizons is not shown.}
\label{fig:Embedding} 
\end{figure}

\section{Changing Spacetime Topology}\label{sec:Topology}
We now turn to the question of whether the double-wormhole geometry of \Fig{fig:CloningWormholes} suffers from any inconsistencies in general relativity. Throughout, we assume that the Einstein equations hold and that the spacetime can be well described by a semiclassical geometry (which corresponds to a choice of how Bob implements the cloning).

The simplest interpretation of the geometry $M$ in \Fig{fig:Embedding} is that, since horizon pairs $AB$ and $AB^\prime$ are each in the thermofield double state \eqref{eq:2CFTs}, the geometries of both wormholes are the same. In this case, the geometry after Bob performs the cloning simply consists of two separate sheets, each a copy of the original ER bridge, glued together along horizon $A$. Note that in this case $M$ contains bifurcate geodesics: any timelike geodesic intersecting horizon $A$ after the cloning occurs will split into two timelike geodesics, one going along the sheet containing $B$ and the other along the sheet containing $B'$. These timelike bifurcate curves indicate a breakdown of the Hausdorff condition, the requirement that for any two points $x\neq y$, there exist disjoint open sets $X\ni x$ and $Y\ni y$.\footnote{Bifurcating geodesics imply failure of the Hausdorff condition, but the converse is not necessarily true; see, for example, the discussion of Taub-NUT space in  Refs.~\cite{HawkingEllis,Hajicek}.} Since the bifurcate timelike curve in question has bounded (being a geodesic, zero) acceleration and moreover the non-Hausdorff boundary of $M$ (horizon $A$) is codimension 1, it follows by a theorem of H{\aa}j{\'i}\v{c}ek \cite{Hajicek} that $M$ is not strongly causal. Strong causality is the requirement that for all points $p\in M$ there is an open neighborhood $P\ni p$ such that any timelike curve passing through $P$ does so only once; this is a weaker condition than global hyperbolicity, so the setup depicted in \Fig{fig:Embedding} leads, via H{\aa}j{\'i}\v{c}ek's theorem, to breakdown of Cauchy evolution \cite{Domain}. Intuitively, this happens because once a timelike curve intersects horizon $A$ it becomes impossible to predict its future. If we wish to avoid immediately abandoning strong causality, we must relax the assumption that the geometry after cloning is merely a two-sheeted copy of the original ER bridge and instead turn to the question of whether the topology change induced by cloning is alone sufficient to guarantee a pathology for a spacetime that remains Hausdorff.

The topology change in question occurs in a localized region of spacetime.  Let us define a partial Cauchy surface \cite{Tipler} to be a spacelike slice through the entire spacetime such that any causal (timelike or null) curve intersects the surface at most once. A 3-surface $\Sigma$ is called externally Euclidean if there exists compact $\Gamma\subset\Sigma$ such that $\Sigma-\Gamma$ is diffeomorphic to Euclidean space minus a 3-ball, i.e., $\Sigma-\Gamma \simeq S^2 \otimes \mathbb{R}$. Given these definitions, we can draw two disjoint externally Euclidean partial Cauchy surfaces $\Sigma_1$ and $\Sigma_2$, where $\Sigma_1$ passes through horizons $A$ and $B$ before the cloning and $\Sigma_2$ passes through horizons $A$, $B$, and $B'$ after the cloning, as shown in \Fig{fig:Embedding}. Importantly, $\Sigma_1$ and $\Sigma_2$ are {\it not} diffeomorphic, $\Sigma_1 \not\simeq \Sigma_2$. Taking $A$, $B$, and $B'$ to be centered on a line on $\Sigma_2$ and quotienting by the rotation group $SO(2)$ around this line, $\Sigma_1/SO(2)$ and $\Sigma_2/SO(2)$ are 2-manifolds with genera 1 and 2, respectively, and are therefore not topologically equivalent. 
The four-dimensional spacetime region whose boundary is $\Sigma_1 \cup \Sigma_2$, called $M$ in \Fig{fig:Embedding}, is externally Lorentzian: there exists a compact manifold $K$ such that $M-K \simeq S^2\otimes\mathbb{R}\otimes[0,1]$, a timelike foliation of spacelike slices $S^2\otimes\mathbb{R}$. Then Geroch's topology-conservation theorem \cite{GerochThesis,Centenary,Tipler} implies that, since $\Sigma_1\not\simeq\Sigma_2$, $M$ must contain a CTC.

While the existence of a CTC somewhere in spacetime is already problematic, we can state a stronger result. We note that $\Sigma_1$ is a Cauchy surface for $M-K$, that is, for all $p \in M-K$, every future- and past-inextendible causal curve through $p$ intersects $\Sigma_1$. Let us assume the generic condition, which asserts that every causal geodesic with tangent vector $k^\mu$ passes through some point for which
\be 
k^\alpha k^\beta k_{[\mu} R_{\nu]\alpha\beta[\rho}k_{\sigma]}\neq 0.\label{eq:generic}
\ee
This means that every timelike or null geodesic experiences a tidal force at some point.\footnote{If the spacetime under consideration has some special symmetry allowing \Eq{eq:generic} to fail for some geodesic, we can enforce the generic condition by simply adding gravitational waves (that is, nonzero Weyl tensor) sufficiently weak to avoid nonnegligible back-reaction on the rest of our argument.} Then Tipler's topology-conservation theorem \cite{Tipler,Tipler2} implies that since $\Sigma_1\not\simeq\Sigma_2$, the NEC\footnote{While \Ref{Tipler} states the theorem in terms of the weak energy condition, this can be strengthened to the NEC as stated in \Ref{Tipler2}.} must fail. That is, the topology change dual to cloning under ER=EPR implies that there must exist fields in the theory for which one can arrange an energy-momentum tensor $T_{\mu\nu}$ such that
\be
T_{\mu\nu} k^\mu k^\nu < 0 \label{eq:NECviolation}
\ee
along some null vector $k^\mu$.

Although violations of the NEC (see also \Ref{Rubakov}) have been shown to occur at a quantum level \cite{Roman}, it has not been shown that such violation is sufficient to allow unusual semiclassical gravitational behavior \cite{MTY,ER=EPR}. However, the NEC violation in the present thought experiment implies macroscopic topology change that results from Bob's cloning procedure with, for example, astrophysical-scale entangled black holes. We conclude that violation of the no-cloning theorem is dual under ER=EPR to topology change and problems with causality, leading to CTCs (by Geroch's theorem) or strong violation of the NEC (by Tipler's theorem).

It is worth noting that the topology theorems do not rule out sensible processes like black hole pair production in the context of ER=EPR. If we consider entanglement as a conserved quantity \cite{Horodecki}, then creation of a pair of entangled black holes does not change the topology, as the ER bridge between them is formed in ER=EPR from the Planckian wormholes connecting the entangled vacuum. Moreover, the process of black hole pair creation is not well described semiclassically, so our results do not apply in that case; in contrast, the cloning process examined in this work can be treated in the setting of semiclassical geometry. Unlike pair production, cloning {\it does} violate the axioms of the topology-conservation theorems precisely because it involves non-unitarily creating entanglement (and therefore wormholes) that did not previously exist.

\section{Wormholes and Causality}\label{sec:Wormholes}
We have shown that violation of the no-cloning theorem is dual under ER=EPR either to immediate breakdown of Cauchy evolution or to severe violation of the NEC [\Eq{eq:NECviolation}]. The latter implies the condition that allows for stabilization of wormholes; specifically, one must have violation of the \textit{averaged} NEC \cite{Kip,MTY}. That is, a traversable ER bridge requires
\be
\int_0^\infty T_{\mu\nu}k^\mu k^\nu \mathrm{d}\lambda<0\label{eq:ANEC}
\ee
for some null geodesics with affine parameter $\lambda$ and tangent vector $k^\mu$. \Ref{MTY} exhibits a construction of a traversable ER bridge that just satisfies \Eq{eq:ANEC} within the wormhole while retaining nonnegative total energy.

The connection between wormhole stabilization and the NEC is highly relevant in the context of the ER=EPR correspondence, as the argument in \Ref{ER=EPR} regarding the impossibility of using wormholes (and by duality, entanglement) to transmit information is critically dependent on the ER bridges pinching off too quickly to allow for signal traversal \cite{Fuller}; a stabilized wormhole would falsify this line of reasoning. Said another way, violation of the NEC plus the existence of wormholes leads to traversable wormholes, which would lead to causality violation. In particular, given a traversable ER bridge, one can immediately form a causal paradox (i.e., a closed signal trajectory) by simply moving the wormhole mouths far apart and giving them a small relative boost \cite{MTY,IRWGC}. The connection between topology change and causality violation in the gravitational sector is now explicit and is satisfyingly analogous to the connection between unitarity/no-cloning and causality on the quantum mechanical side of the ER=EPR duality.

\section{Perspectives for Future Work}\label{sec:Conclusions}
As we have seen, spacetime topology change leads inexorably to violation of causality, via either breakdown of the Hausdorff condition or creation of traversable wormholes. Using ER=EPR to translate this result to quantum mechanics, we find that violation of the axioms of the topology-conservation theorems is dual to violation of monogamy of entanglement (i.e., cloning) and the existence of wormholes is dual to the existence of entanglement entropy. The logical flow of our reasoning is:
\be 
\begin{aligned} & \begin{aligned}\text{{C}} & & \& & & \exists\;\text{QE} &  & \implies & & \text{SLS}\\
\updownarrow & & & & \updownarrow & & & & \updownarrow\\
\Delta\text{{T}} & & \& & & \exists\;\text{WH} &  & \stackrel{{\cancel{\text{{NEC}}}}}{\implies} &  & \text{TWH}\\
\Downarrow
\end{aligned}
\\
 & \big(\cancel{\text{{NEC}}}\;\&\;\exists\;\text{{CTCs}}\big)\;||\;\cancel{\text{{SC}}}
\end{aligned}\label{eq:logic}
\ee
Here, C denotes ``quantum cloning'', ``QE'' quantum entanglement, ``SLS'' superluminal signaling, ``T'' topology, ``WH'' wormholes, ``TWH'' traversable wormholes, and ``SC'' strong causality. The single-lined arrows in \Eq{eq:logic} indicate duality of specific statements under ER=EPR, double-lined arrows indicate logical implication, and strikethroughs indicate violation.

It is striking that on both the general relativistic and quantum mechanical sides of the duality, violation of the no-go theorem leads to problems for causality. The unexpected connection between cloning and topology change offers support for the ER=EPR correspondence, which provides a natural explanation for their relation.

A promising avenue for future research is the investigation of whether other no-go theorems in quantum mechanics and gravity neatly correspond under ER=EPR. The no-deleting theorem corresponds to the topology theorem in exactly the same way as the no-cloning theorem, while the no-communication theorem is equivalent to the assertion of nontraversability of wormholes. On the gravity side, violation of Hawking's area theorem, i.e., the generalized second law of thermodynamics, requires either breakdown of cosmic censorship or of the null energy condition \cite{Frolov}, the latter allowing wormhole traversal \cite{MTY}. In ER=EPR, this corresponds to violation of the no-communication theorem \cite{ER=EPR} and, in AdS/CFT, would correspond to violation of unitarity in the dual CFT state of \Eq{eq:2CFTs} \cite{IRWGC}. Whether all known gravitational or quantum mechanical no-go theorems map onto each other in this way is a fascinating open question. More generally, the connections among infrared constraints on ultraviolet physics, such as unitarity and causality \cite{IRWGC,IRUV,2ndLaw,Brun}, will continue to play an important role in understanding quantum gravity.

\noindent {\it Acknowledgments}: We thank Sean Carroll, Clifford Cheung, Shamit Kachru, Stefan Leichenauer, and John Preskill for helpful discussions.
This research was supported in part by DOE grant DE-SC0011632 and by the Gordon and Betty Moore Foundation through Grant 776 to the Caltech Moore Center for Theoretical Cosmology and Physics. N.B. is supported by the DuBridge postdoctoral fellowship at the Walter Burke Institute for Theoretical Physics. G.N.R. is supported by a Hertz Graduate Fellowship and a NSF Graduate Research Fellowship under Grant No.~DGE-1144469.

\bibliographystyle{utphys}

\bibliography{EPRbib}

\end{document}